\documentclass[10pt,letterpaper]{article}

\usepackage{cogsci}

\cogscifinalcopy 

\usepackage{pslatex}
\usepackage{apacite}
\usepackage{float} 
\usepackage{amsmath} 
\usepackage{graphicx} 
\usepackage{booktabs} 
\usepackage{cuted} 
\usepackage{etoolbox} 
\newcommand{\tabitem}{~~\llap{\textbullet}~~}

\title{Complementary Structure-Learning Neural Networks for Relational Reasoning}
 
\renewcommand{\thefootnote}{\fnsymbol{footnote}} 
  
\author{{\large \bf Jacob Russin$^{1}\thanks{Equal contribution}$ (jlrussin@ucdavis.edu)} \\
  \And {\large \bf Maryam Zolfaghar$^{2}\footnotemark[\value{footnote}]$ (mzolfaghar@ucdavis.edu)} \\
  \AND {\large \bf Seongmin A. Park$^3$ (apark@ucdavis.edu)} \\
  \And {\large \bf Erie Boorman$^{1,3}$ (edboorman@ucdavis.edu)} \\
  \AND {\large \bf Randall C. O'Reilly$^{1,2,3}$ (oreilly@ucdavis.edu)} \\
  }

\begin{document}

\maketitle

\begin{strip}
{\centering
    $^1$ Department of Psychology, UC Davis; $^2$ Department of Computer Science, UC Davis;  
    
    $^3$ Center for Mind and Brain, UC Davis;  $^4$ Center for Neuroscience, UC Davis
\par}
\end{strip}

\renewcommand{\thefootnote}{\arabic{footnote}}  

\begin{abstract}
The neural mechanisms supporting flexible relational inferences, especially in novel situations, are a major focus of current research.  
In the complementary learning systems framework, pattern separation in the hippocampus allows rapid learning in novel environments, while slower learning in neocortex accumulates small weight changes to extract systematic structure from well-learned environments. In this work, we adapt this framework to a task from a recent fMRI experiment where novel transitive inferences must be made according to implicit relational structure. We show that computational models capturing the basic cognitive properties of these two systems can explain relational transitive inferences in both familiar and novel environments, and reproduce key phenomena observed in the fMRI experiment. 

\textbf{Keywords:} 
neural networks; cognitive maps; complementary learning systems; structure learning; transitive inference
\end{abstract}

\section{Introduction}

Humans and non-human animals are capable of navigating efficiently in both novel and familiar environments. 
For example, in a well-learned environment like one's hometown, it is easy to navigate to new goal locations and plan novel routes. 
When traveling in a new city, it is also possible to navigate to a novel location by reasoning over recent experiences --- even those accumulated on the same day. 
In both cases, efficiency requires processes or representations that allow generalization beyond previous experience.
This kind of generalization has been a long-standing issue in cognitive science, and was integral to early arguments against behaviorism, where it was claimed that a simple stimulus-response mapping could not account for such behaviors \cite{Tolman48}.

More recent work has investigated the computational and neural mechanisms underlying cognitive maps, or representations that capture the structure of the environment and thereby support generalization \cite{ParkMillerNiliEtAl20, WhittingtonMullerMarkEtAl20, BehrensMullerWhittingtonEtAl18a}.
This work has emphasized the importance of certain neocortical areas such as the entorhinal cortex (EC) for spatial reasoning and vector-based navigation \cite{MoserKropffMoser08a}. 
Furthermore, it has been argued that these structured spatial representations may be leveraged for other kinds of abstract relational reasoning in humans \cite{BehrensMullerWhittingtonEtAl18a}.  
Relatedly, although neural networks have enjoyed massive success on difficult machine-learning tasks in recent years
these models are known to fail on out-of-distribution or extrapolation problems \cite{LakeUllmanTenenbaumEtAl17} such as those requiring transitive inferences. 

Here, we apply the well-supported complementary learning systems (CLS) framework \cite{McClellandMcNaughtonOReilly95,OReillyBhattacharyyaHowardEtAl11} to explore two qualitatively different neural mechanisms underlying spatially-grounded relational reasoning abilities in novel and familiar environments.
The CLS framework has emphasized the computational justification for learning mechanisms unfolding on two different timescales, as supported by separate brain areas. 
Slow learning in neocortex allows for the development of more abstract representations that integrate across many experiences and can be leveraged to make novel inferences. 
However, this kind of learning is not possible in naturalistic environments where sequences of events are
not presented in an interleaved or random order, as when one explores only one part of an environment at a time.
This is due to the well-known \textit{catastrophic forgetting} phenomenon, where previous learning is erased by new experiences when learning occurs too quickly or training is not sufficiently interleaved \cite{McClellandMcNaughtonOReilly95}.
The CLS framework proposes that fast learning can occur in the hippocampus due to its pattern-separated, sparse representations. 
These representations have little overlap across examples, and therefore allow fast learning of novel episodes, i.e., \textit{episodic memory} \cite{YonelinasRanganathEkstromEtAl19}, to occur without catastrophic interference.

In the CLS framework, slow cortical learning is needed to build up structural or relational representations over time, which provide the foundation for systematic inferences.  
However, for more unfamiliar situations, rapid hippocampal learning is required.
Previous work has found evidence suggesting a role for the hippocampus in rapid generalization \cite{Eichenbaum04a, ZeithamovaSchlichtingPreston12}, and that a hippocampal model informed by the CLS framework can explain these findings when it is augmented with a recurrent similarity-based computation, proposed to be supported by ``big-loop'' recurrence between the hippocampus and the neocortex and within the hippocampus itself \cite{KumaranMcClelland12a}. 

Here, we build on this work and investigate the interplay between slow generalization in neocortex and rapid generalization in the episodic memory system with computational models based on the principles of the CLS framework. 
Our model of the episodic memory system is similar to previous work \cite{KumaranMcClelland12a} in that it allows rapid generalization in unfamiliar environments, but relies on different computational mechanisms to do so (see Discussion). 
Our models of the cortical system and the episodic memory system were both tested on a novel non-spatial structure-learning paradigm from a recent fMRI experiment \cite{ParkMillerNiliEtAl20}.
Importantly, the task required transitive inferences based on learning over two different timescales: training experience over multiple days, and training examples given on the same day as the inference test. 
In the following, we briefly outline the key findings of the experiment and offer a conceptual framework that integrates them with the CLS perspective. 
We then describe the computational models that were built to capture the basic properties of the proposed conceptual framework, and show that these models are capable of performing transitive inferences in the same task and reproduce other key findings from \citeA{ParkMillerNiliEtAl20}. 

\begin{figure*}[h]
  \centering
  \includegraphics[width=0.65\textwidth]{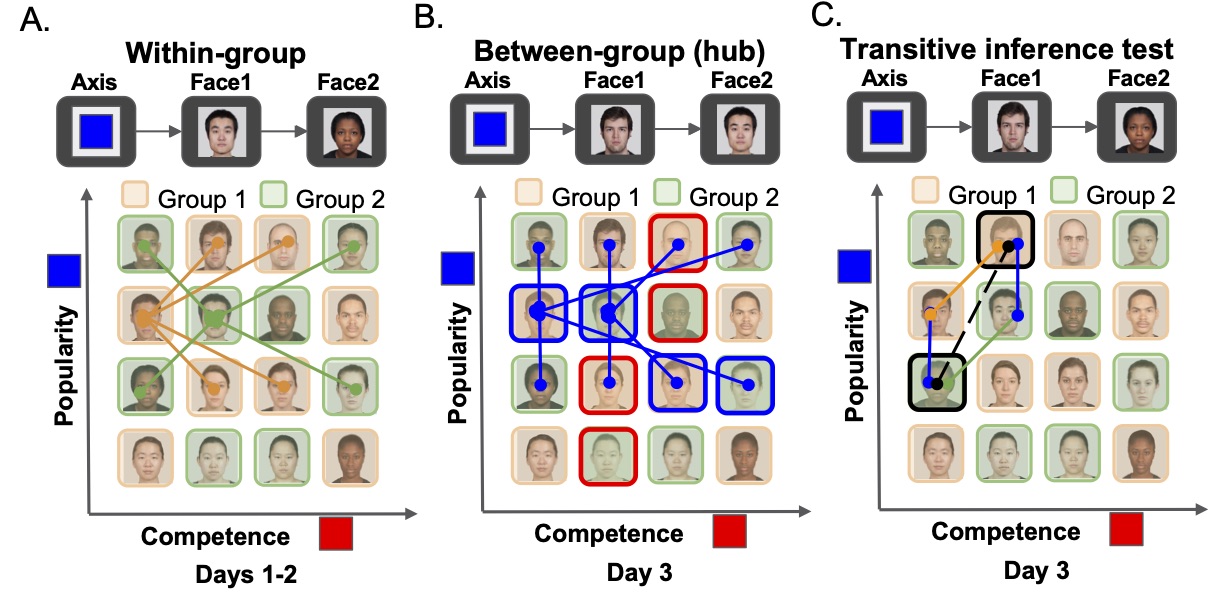}
  \caption{Experimental paradigm used in \citeA{ParkMillerNiliEtAl20}. Participants learned the relative ranks of pairs of faces on an implicit grid with two axes: competence and popularity. These faces were split into two groups (shown in green and orange). \textbf{A)} Over the first two days, participants were trained on within-group pairs that differed by a rank of 1 along each designated axis. Example pairs on the popularity axis are shown with orange lines (within group 1) and green lines (within group 2). \textbf{B)} On the third day, participants learned between-group pairs containing exactly one hub linking the two groups. There were a total of 8 hubs, and each was associated with a certain axis (shown by red and blue outlines). Each hub was paired with the 4 faces from the other group that differed by a rank of 1 along the designated axis. Examples of such pairs are shown for two hubs with blue lines (indicating the popularity axis). \textbf{C)} The third day included the fMRI experiment and transitive inference test. Participants were tested on pairs of faces from different groups that could be connected through one of the two hubs on the appropriate axis. Green, orange, and blue lines indicate the training pairs (2 within-group and 2 between-group, which are shown in A and B) that could be used to make the transitive inference for the pair indicated by the black dotted line.}
\label{fig:task}
\end{figure*}

\subsection{fMRI Experiment}

\citeA{ParkMillerNiliEtAl20} studied the neural mechanisms underlying transitive inference performance on the structure-learning task illustrated in Figure \ref{fig:task}. 
Participants learned to make judgments about the ``popularity'' or ``competence'' of 16 people through pair-wise comparisons along one of these two axes at a time. 
Unknown to the participants, these 16 faces were arranged in a 4x4 2D grid, and were implicitly separated into two groups.
In the first two days of training participants only learned about within-group pairs that different by a rank of 1 (see Figure \ref{fig:task}A). 
On the third day of the experiment, participants learned about between-group pairs containing certain faces that acted as hubs between the two groups (see Figure \ref{fig:task}B). 
This training provided sufficient evidence to allow participants to integrate their previously separated cognitive maps, but was conducted on the same day as fMRI scanning.
In the scanner, participants performed a transitive inference test in which unseen pairs of faces from different groups were compared (see Figure \ref{fig:task}C).
For each of these test pairs, one of two corresponding hubs could be used to make the transitive inference.
The results we focused on in our work can be summarized as follows:
\begin{enumerate}
\itemsep0cm 
    \item Participants exhibited good transitive inference performance, achieving 93.6\% mean accuracy on the unseen pairs tested in day 3. 
    \item Map-like representations were found in several brain areas, including ventromedial prefrontal cortex (vmPFC) and entorhinal cortex (EC). Patterns of activity in these areas demonstrated sensitivity to the ground-truth Euclidean distances between faces in the implicit grid. However, these effects were significantly reduced when the analysis was restricted to between-group pairs that were not encountered during training. 
    \item A repetition-suppression analysis in hippocampus suggested that one of the two relevant hubs was retrieved from episodic memory at the time of inference. 
\end{enumerate}

Taken together, these findings suggest that cortical learning systems in vmPFC and EC were able to integrate across the pairs of faces encountered during training to form map-like representations that would be useful for making transitive inferences within groups. 
However, the effects in these areas were reduced when the analysis was restricted to novel between-group pairs, and participants seemed to retrieve the relevant hubs from episodic memory in hippocampus during the transitive inference test. 
Thus, although the within-group pairs were well-learned over the first two days of training, these groups may not have been fully integrated into a single coherent cognitive map at the time of testing.
This may have forced participants to rely instead on hippocampal retrieval of recently-learned between-group training episodes (which always included a hub) to generalize during the transitive inference test.
Thus, there appear to be two separable cognitive mechanisms that allow for relational transitive inferences to be made in this task: 1) if given enough training time, cortical areas such as vmPFC and EC can learn representations that reflect the implicit relational structure of the grid, and 2) an episodic retrieval mechanism can ensure good transitive inference performance with pairs that were seen only on the same day as the test. 
Below we outline a general framework that integrates these findings, and the apparent redundancy in these two systems, with the CLS perspective. 

\section{Complementary Structure-Learning Systems}

The CLS framework explains how the brain can support integrative representation learning without suffering from catastrophic forgetting \cite{McClellandMcNaughtonOReilly95, OReillyBhattacharyyaHowardEtAl11}. 
However, the CLS framework also emphasizes other important reasons for fast learning in an episodic memory system. 
In particular, slow cortical learning may be insufficient to allow for efficient adaptation in relatively unfamiliar environments \cite{KumaranMcClelland12a}. 
The findings from \citeA{ParkMillerNiliEtAl20} suggest that humans are capable of making novel transitive inferences using experiences acquired on the same day.
Furthermore, they show that these inferences are mediated by hippocampal retrieval of the intermediate states (i.e., hubs) that would allow such inferences to occur. 
Taken together, these findings suggest that the dual-process view emphasized in CLS may explain the apparent redundancy in structure-learning mechanisms studied in neuroscience and psychology (see Table \ref{tab:tradeoffs}). 

\begin{table}[h]
\begin{center} 
\caption{Complementary structure-learning systems.} 
\label{tab:tradeoffs} 
\vskip 0.12in
\begin{tabular}{l|p{5.1cm}} 
\hline
System  &  Properties \\
\hline
Cortical learning &  \tabitem Learns slowly through small, incremental weight changes \\
                  &  \tabitem Inference is fast and less effortful with map-like representations \\
\hline
Episodic memory   &  \tabitem Learning can be fast due to sparse, pattern-separated representations \\
                  &  \tabitem Inference is slower, requiring cognitive control for deliberate, goal-directed retrieval \\
\hline
\end{tabular} 
\end{center} 
\end{table}

In the case of spatial navigation, slow cortical learning can integrate across many experiences to form map-like representations. 
This system is capable of directly utilizing its integrative representations without further processing, and can thus make inferences rapidly.
However, this system would not be able to make inferences in a newly learned environment if it did not have time to integrate across particular episodes \cite{KumaranMcClelland12a}.
This may have been the case in the transitive inference test conducted on the same day as the between-group training in the fMRI study \cite{ParkMillerNiliEtAl20}. 
Fast episodic learning, on the other hand, can immediately store memories of individual experiences, allowing inferences to be made in unfamiliar environments based on few such experiences. 
However, the episodic nature of its representations do not allow the sort of direct inferences that are available to the cortical system. 
Instead, transitive inferences require a slower, more deliberate process of goal-directed retrieval and further processing of the stored memories \cite{ZeithamovaSchlichtingPreston12}. 
An organism equipped with both systems would be capable of making novel inferences in both familiar and unfamiliar environments. 
In the following, we provide evidence from models that capture, on a \emph{computational} level, the basic properties of the proposed complementary structure-learning systems, and show that these systems reproduce key findings from \citeA{ParkMillerNiliEtAl20}.

\section{Modeling Framework}

We simulated\footnote{All data and code used for experiments and analyses are available at https://github.com/MaryZolfaghar/CSLS} each of our models on the training and testing procedure used in the task, including its within-group and between-group structure and transitive-inference test. 
In particular, each trial consisted of a presentation of two faces and the axis along which the judgment should be made (i.e., ``competence'' or ``popularity''). 
The models were required to make a binary judgment about whether the first face ranked higher or lower than the second face along the specified axis.

\begin{figure}[h]
  \includegraphics[width=0.5\textwidth]{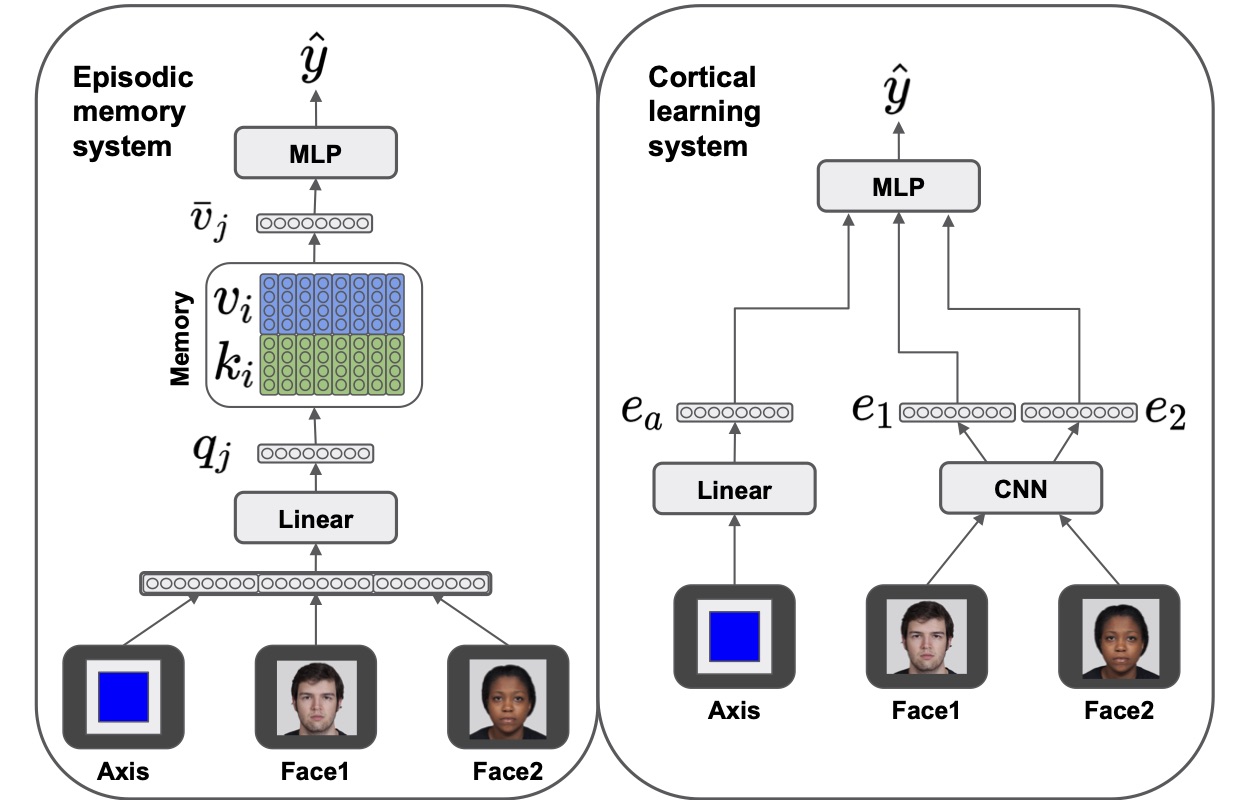}
  \caption{Model architecture. (Left) The episodic memory system stores representations of individual training trials in a key-value memory. New inferences are made by querying the memory to retrieve the relevant trials, which are then processed by an MLP to generate an answer. (Right) The cortical learning system was modeled as a simple feedforward network with convolutional layers to process the images. This system relies on its learned representations to perform transitive inferences. }
\label{fig:model}
\end{figure}

\subsection{Cortical Map-Building}

The cortical representation-learning system should accumulate small updates over many trials to build map-like representations that can be directly utilized to make transitive inferences. 
We modeled this process with a simple feedforward neural network with two convolutional layers (see right side of Figure \ref{fig:model}). 
Face images were taken from the same database used in the fMRI experiment \cite{StrohmingerGrayChitucEtAl16}, and were downsampled to 64x64 and grayscaled for faster simulation. 
The within-group and between-group hub samples were all trained simultaneously (i.e., the pairs that were trained on different days of the fMRI experiment were trained simultaneously in the model).
This is because the purpose of our model of the cortical system was to show that, if given enough training time, it could perform transitive inferences based on its learned representations, and allow fast inference in familiar environments. 
Each face was processed with the same convolutional neural network, and the axis variable, encoded as a one-hot vector, was embedded with a linear layer.
These three embeddings were then concatenated and passed through a multi-layer perceptron (MLP) with rectified linear unit (ReLU) activation functions.
This network captures the basic properties of slow cortical learning in that it accumulates small updates to its synaptic weights over many trials, and makes inferences directly based on its learned representations of each face. 

\subsection{Goal-Directed Episodic Memory Retrieval}

The episodic memory system should learn quickly by storing individual training episodes, and make inferences by retrieving the previous trials that are relevant to the current one \cite{McClellandMcNaughtonOReilly95, KumaranMcClelland12a}. 
For this purpose, we used a neural memory system (see left side of Figure \ref{fig:model}) with a soft retrieval mechanism \cite{BotvinickRitterWangEtAl19a}. 
This memory system immediately stores each trial ($x_i$) seen during training as a key, value pair: $ k_i = W_k x_i$, $v_i = W_v x_i $, where $k_i$ is the key, $v_i$ is the value, and $x_i$ is the trial, which is a concatenation of a one-hot encoding of each face, the axis variable ($a$), and the correct answer ($y$) of the $i$th trial. 
One-hot encodings were used for faces under the assumption that what is stored in the episodic memory system should be a highly processed, sparse encoding \cite{McClellandMcNaughtonOReilly95}. 
To make an inference, the model generates a query according to the current pair of faces: $ q_j = W_q x_j^- + b_q $, where $x_j^-$ indicates the $j$th test trial with the same components but excludes the correct answer ($y$). 
This query is then used to retrieve the memories most relevant to the current trial:
\begin{equation}
    \bar{v}_j = \text{softmax} \left( q_j K^T \right) V
    \label{eq:retrieval}
\end{equation}
where $K$ and $V$ are matrices containing all of the stored memories. 
Finally, the retrieved memories $\bar{v}_j$ are passed through an MLP to produce the final answer: $ \hat{y_j} = \text{MLP}(\bar{v}_j)$.
This network captures the basic properties of a fast-learning episodic memory system in that each training episode can be stored in memory immediately upon presentation, and must later be retrieved in a goal-directed way to make a transitive inference. 

An interesting problem in modeling episodic memory concerns the learning mechanisms involved in goal-directed memory retrieval. 
We assume that the human participants recruited for the \citeA{ParkMillerNiliEtAl20} study had extensive prior experience with goal-directed memory retrieval and everyday transitive inferences. 
We therefore adopted a meta-learning strategy \cite{SantoroBartunovBotvinickEtAl} to model this prior experience, and pretrained the episodic memory system to learn to solve new transitive inference problems sampled from a distribution of such tasks. 
This pretraining consisted of slow, incremental changes to the weights responsible for mapping into and out of the episodic memory itself, and should thus be thought of as occurring in memory-related cortical areas rather than in the hippocampus proper \cite{McClellandMcNaughtonOReilly95}. 
The system was pretrained on a distribution that was generated by permuting the positions of each face in the 4x4 grid.
For each new task, the memory system stored training samples in its memory and used them to make transitive inferences in the testing phase, where it accumulated errors that were then used to update its learnable parameters.
The model was then tested on how well it could generalize with a new configuration of faces it had never seen before.

This kind of meta-learning strategy was adopted from previous work \cite{Lake19}, and shares with it the limitation that the pretraining tasks are unrealistically similar to the final test --- future work will examine the extent to which the model can generalize when trained on substantially different goal-directed retrieval and transitive inference tasks. 
Additionally, although the resulting goal-directed retrieval mechanism in this model does not capture the hypothesized properties of being deliberative and requiring cognitive control (thus making inferences slower), a more biologically grounded approach involving frontal cortical executive function systems, planned for future work, would do so.
Our purpose in the current study was to show that this system was capable of making transitive inferences in a structured environment.

\subsection{Implementation Details}

Models were built using PyTorch. 
Models were trained with a cross-entropy loss function and Adam optimizer \cite{KingmaBa15} with a batch size of 32 and a learning rate of 0.001.\footnote[1]{Note that the ``learning rate'' for the episodic memory refers to the weight updates in the pre-training phase. During training, it immediately stored experiences upon presentation.}
The cortical system was trained for 100 epochs with a batch size of 32. 
The axis embedding ($e_a$) had 32 dimensions. 
Convolutional layers had no padding, a kernel size of 3, a stride of 2, and 4 and 8 channels in the first and second layers, respectively.
Each convolutional layer was followed by a max-pooling layer with a kernel size of 2. 
The CNN contained a linear layer to produce flat 72-dimensional vectors $e_1$ and $e_2$, which were passed to the final MLP, which had 128 hidden units. 
The episodic memory system was pre-trained on 10,000 permutations. 
Queries, keys, and values were all 32-dimensional, and the final MLP had 64 hidden units. 

\section{Results}

Both systems proved to be capable of performing transitive inferences in the task environment from \citeA{ParkMillerNiliEtAl20}: each system achieved 100\% accuracy on the held-out test set in which unseen between-group pairs were tested. 
This validates the idea that the two qualitatively different kinds of learning system outlined above are capable of reproducing human transitive inference performance on the task.
To investigate how these qualitative differences might have affected each model's inference strategy, we performed analogues of key analyses done in the experiment \cite{ParkMillerNiliEtAl20} to interpret the behavior of each system, and to evaluate them against empirical results obtained in the fMRI experiment.

\subsection{Cortical Representations Reflect Task Structure}

\begin{figure}[h]
  \includegraphics[width=0.5\textwidth]{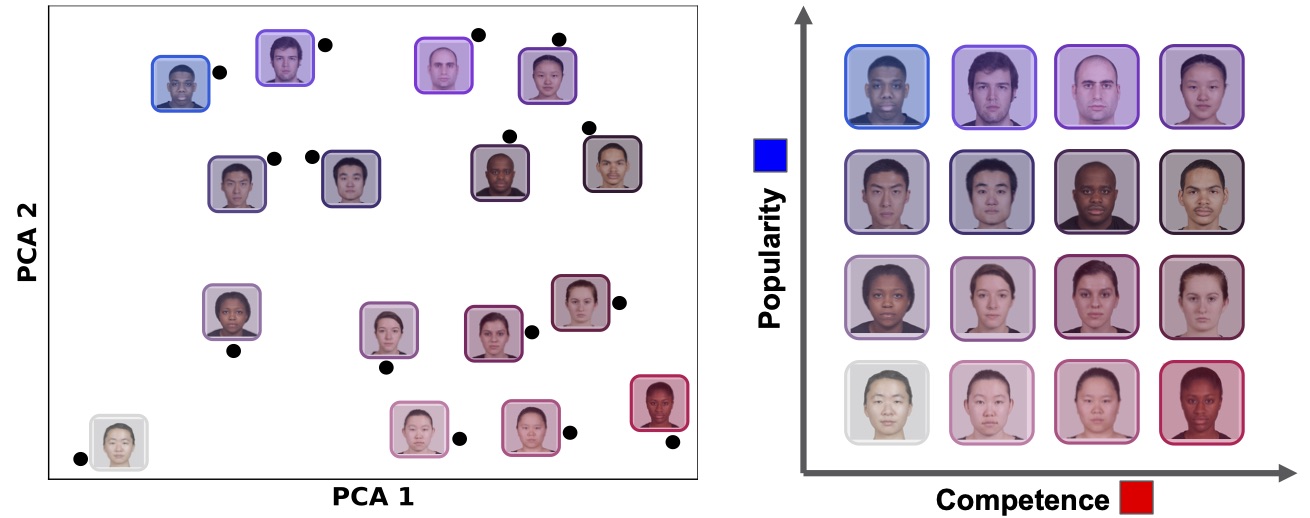}
  \caption{Visualization of embeddings learned by the cortical system. Embeddings for each face was projected into two dimensions using PCA, and then rotated by a fixed angle for illustration purposes. The relative positions of the representations indicate that the model has learned to represent the faces in terms of their implicit relational structure.}
\label{fig:pca}
\end{figure}

To understand how the cortical system had learned to represent each of the faces, we conducted analyses on the embeddings of each face obtained from the CNN.
Visualization of these embeddings with principal components analysis (PCA; see Figure \ref{fig:pca}) showed that the cortical system had learned to represent the faces in terms of their structured relationships, i.e., it had learned map-like representations. 
These top two principal components explained 95.1 \% of the variance in the embeddings, indicating that the model had learned to represent the faces on a near two-dimensional grid. 

In addition to the PCA, we conducted an analysis similar to those done in the fMRI experiment \cite{ParkMillerNiliEtAl20}, where patterns of activity in vmPFC and EC were found to be sensitive to Euclidean distances in the ground-truth grid.
We measured the Pearson correlation between ground-truth Euclidean distances in the grid and the observed distances between each pair of embeddings. 
A strong correlation was observed ($r(118) = .910$, $p < 0.001$), indicating the same sensitivity to structured relationships in the grid. 

\subsection{Episodic Memory System Retrieves Hubs}
\begin{figure}[h]
  \includegraphics[width=0.5\textwidth]{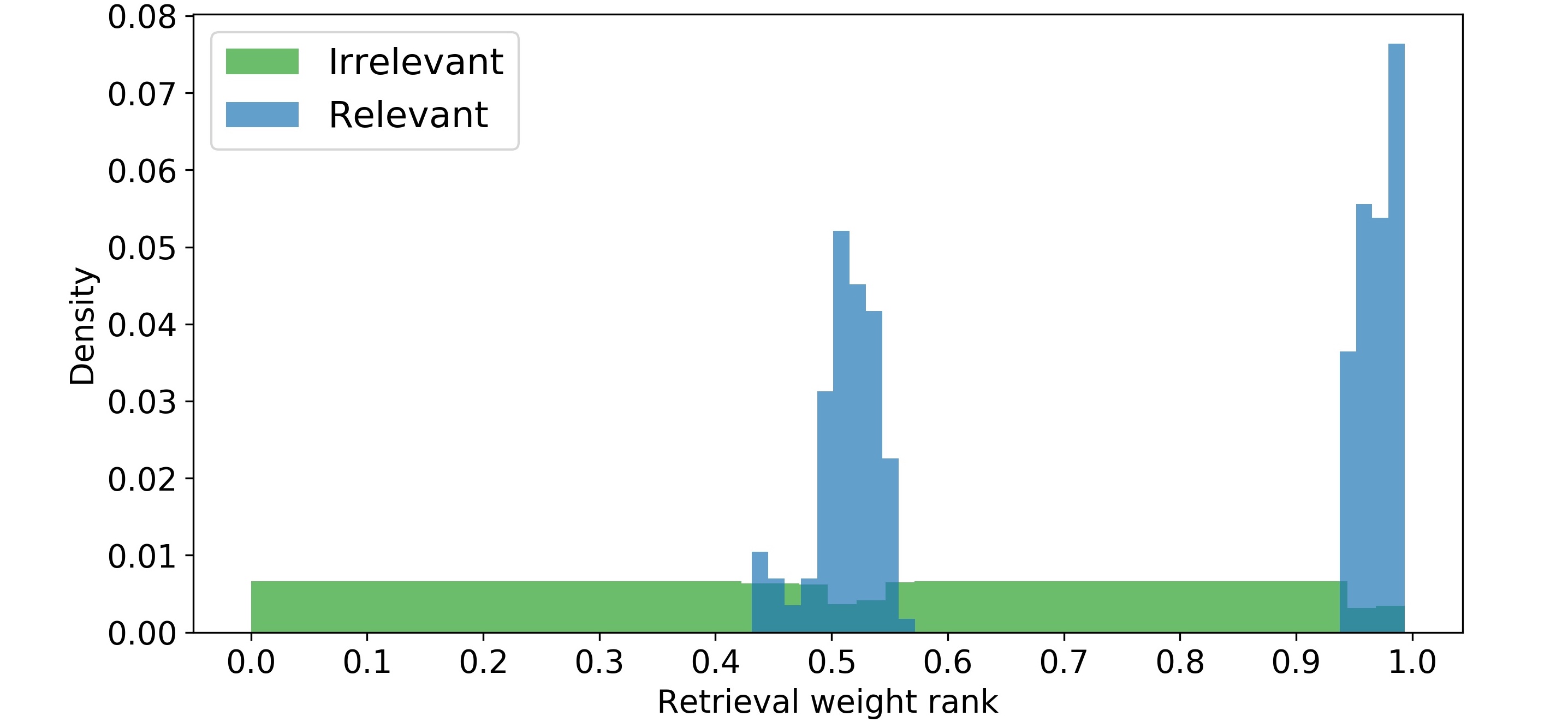}
  \caption{Histograms of relevant and irrelevant trials retrieved from the episodic memory during testing. Relevant memories, which always contained a hub, made up the majority of those with the highest weights. This reproduces the fMRI finding that the hubs were retrieved during the inference test. Note that it was not necessary to retrieve every relevant memory to get the correct answer, which may be why the relevant memories were not always retrieved with the highest weights. Counts were normalized to probability densities.}
\label{fig:retrieval}
\end{figure}

In the original fMRI experiment, a repetition-suppression analysis suggested that participants were retrieving the relevant hubs from hippocampus during the transitive inference test (see Figure \ref{fig:task}C). 
Although the episodic memory model did not have analogous neural adaptation dynamics that would allow us to model repetition suppression, we conducted an analysis on the retrieved memories to see how the hubs were being used to make transitive inferences. 
The soft episodic retrieval mechanism shown in equation (\ref{eq:retrieval}) uses a softmax to produce a probability distribution over all of the items in memory. 
For each test trial, we directly analyzed the weights applied to the memories for the relevant hub trials and compared these weights to the irrelevant memories (see Figure \ref{fig:retrieval}). 
Memories were counted as relevant if they included one of the two possible between-group hubs for the given pair of faces, and connected this hub to one of the two faces from the current trial (see Figure \ref{fig:task}C).
This revealed that the weights applied to the relevant hub memories were usually the largest (i.e., the hub trials were retrieved more than the irrelevant trials). 
Furthermore, an additional analysis found that in every test trial, one of the two possible ``paths'' connecting the first face to the second face (e.g., in Figure \ref{fig:task}C, the path through the blue line and green line or the path through the blue line and orange line) was in the top 5\% of retrieved memories. 

\section{Discussion}

The CLS perspective emphasizes the need for two qualitatively different learning systems in the brain: fast learning can occur in the hippocampus due to its pattern-separated representations, while learning in the neocortex must be slow due to its overlapping representations \cite{McClellandMcNaughtonOReilly95}. 
Here, we investigate this conceptual framework in the domain of structure-learning and relational transitive inference \cite{KumaranMcClelland12a}, and propose an analogous distinction.
The episodic memory system can learn quickly and generalize in relatively unfamiliar environments, but requires a more deliberate goal-directed retrieval process. 
The cortical system learns slowly but can make fast inferences in familiar environments from its learned representations. 
As in the traditional CLS framework, an organism equipped with both systems would retain the benefits of each, allowing generalization in both novel and familiar environments.  
Our computational models provide evidence that each of the two proposed systems are able to perform well on a difficult relational transitive-inference test under different circumstances: the cortical system can make these inferences once extensive experience with an environment has been accumulated, while the episodic system can do so quickly, as long as it has had sufficient prior exposure to similar tasks. 
Our models also reproduce the basic findings from a human fMRI experiment \cite{ParkMillerNiliEtAl20}: the cortical system learns map-like representations that encode the implicit relational structure of the grid, while the episodic memory system learns to query its memory for the appropriate hubs connecting the two groups.

\citeA{KumaranMcClelland12a} investigate rapid generalization in a hippocampal model based on the principles of the CLS framework. 
The model allows retrieval-based inferences to be made --- despite the nature of its pattern-separated representations --- by incorporating a recurrent similarity computation that can perform associative linking \cite{Eichenbaum04a, OReillyRudy01}. 
This computation is hypothesized to be supported by ``big-loop'' recurrence \cite{KosterChadwickChenEtAl18}. 
Our model of the episodic memory system is not inconsistent with hippocampal retrieval-based inferences based on dynamic similarity computation, and in fact the fMRI experiment showed evidence of the presence of such similarity structure in the hippocampus \cite{ParkMillerNiliEtAl20}. 
In addition, the strategy used by our model to solve transitive inference problems appeared consistent with the associative linking exhibited by the model of \citeA{KumaranMcClelland12a}, as shown by the retrieval of hubs linking the two groups (see Figure \ref{fig:retrieval}). 
However, in our model this strategy emerged over the course of (meta-)learning the structure of transitive inference problems, suggesting a more general mechanism that could be applied to goal-directed retrieval tasks that are not solvable with an associative linking strategy. 
This learning mechanism has been shown to be useful in the context of one-shot learning \cite{SantoroBartunovBotvinickEtAl}, and compositional generalization \cite{Lake19}.
More work is needed to investigate whether hippocampal involvement in rapid generalization occurs when such a strategy is not possible, and whether our model would benefit from the recurrent computation intrinsic to the model of \citeA{KumaranMcClelland12a}.

Our modeling framework shares important properties with the Tolman-Eichenbaum Machine (TEM) \cite{WhittingtonMullerMarkEtAl20}, which also incorporates meta-learning and models structure-learning in EC. 
A critical difference between these two models is that in TEM, structure-learning depends on backpropagating error signals through the hippocampus, whereas the CLS framework holds that slow cortical learning can operate independent of the hippocampus to facilitate inferences, consistent with the remarkably intact abilities of early developmental amnesics \cite{Vargha-KhademGadianWatkinsEtAl97}.

Our proposed framework integrates ongoing empirical findings about cognitive maps 
with the CLS perspective, but it also shares some similarities to other prominent dual-process views in cognitive science. 
For example, prominent theories emphasize a distinction between habitual and controlled processing \cite{OReillyNairRussinEtAl20a}, fast and slow thinking \cite{kahneman2011thinking} and model-free and model-based RL \cite{BotvinickRitterWangEtAl19a}. 
Our conceptual framework proposes a similar distinction between the deliberative, goal-directed retrieval that must occur in the episodic memory system to make transitive inferences, and the more automatic or vector-based generalization that can occur in the cortical system in familiar environments. 

There are some important limitations of our current computational models that must be addressed in future work. 
First, although the two proposed cognitive systems are hypothesized to be realized in the hippocampus and cortical areas such as EC, we have not focused on the interactions that should occur between the two systems.
For example, the representations stored in episodic memory should be directly informed by the slowly changing representations learned in cortex, reflecting cortical inputs to the hippocampus.
The fMRI study found that map-like representations were also present in the hippocampus \cite{ParkMillerNiliEtAl20}, perhaps due to interactions with nearby cortical areas \cite{KumaranMcClelland12a}. 
A more integrated model would show how map-like representations in cortex can influence hippocampal processing, and how reliance on the episodic memory early in learning shifts to reliance on the cortical system later in learning. 
This shift may occur due to the cognitive demands imposed on an episodic retrieval mechanism required to reason over individual past experiences.
The current episodic memory system does not capture the cognitive control hypothesized to be required for inferences to be made; future work will address this with a more integrated model that deploys an episodic retrieval mechanism with costly sequential processing. 
Finally, the neural networks used in our models biologically implausible in a number of ways, e.g., the use of a slot-based episodic memory and the standard backpropagation algorithm. 
Future work will focus on more biologically plausible learning algorithms and more detailed biology of the neocortex and hippocampus.

\subsection{Acknowledgments}
We would like to thank the members of the Computational Cognitive Neuroscience lab and the Learning and Decision Making lab, as well as reviewers for helpful comments and discussions. 
The work was supported by: ONR grants ONR N00014-20-1-2578, N00014-19-1-2684 / N00014-18-1-2116, N00014-18-C-2067.
J.R. was supported by the NIMH under Award Number T32MH112507. 
The content is solely the responsibility of the authors and does not necessarily represent the official views of the NIH. 

\bibliographystyle{apacite}

\setlength{\bibleftmargin}{.125in}
\setlength{\bibindent}{-\bibleftmargin}

\bibliography{cogsci2021}

\end{document}